\documentclass[article]{elsarticle}

\usepackage{lineno,hyperref}
\usepackage{amsfonts}
\usepackage{amsmath}
\usepackage{amssymb}
\usepackage{amsbsy}
\usepackage{graphicx}
\usepackage{graphics}

\biboptions{numbers,sort&compress}

\modulolinenumbers[5]
\journal{Annals of physics}
\begin{document}
\begin{frontmatter}

\title{Geometric Curvature and Phase of the Rabi model}
\author[rvt]{Lijun Mao}
\author[rvt]{Sainan Huai}
\author[rvt]{Liping Guo}
\author[rvt]{Yunbo Zhang\corref{cor1}}
\ead{ybzhang@sxu.edu.cn} \cortext[cor1]{Corresponding author}
\address[rvt]{Institute of Theoretical Physics
and Department of Physics, Shanxi University, Taiyuan 030006, China}


\begin{abstract}
We study the geometric curvature and phase of the Rabi
model. Under the rotating-wave approximation (RWA), we apply the gauge
independent Berry curvature over a surface integral to calculate the Berry
phase of the eigenstates for both single and two-qubit systems, which is found
to be identical with the system of spin-1/2 particle in a magnetic field.
We extend the idea to define a vacuum-induced geometric curvature
when the system starts from an initial state with pure vacuum bosonic field.
The induced geometric phase is related to the average photon number in a period
which is possible to measure in the qubit-cavity system. We also calculate
the geometric phase beyond the RWA and find an anomalous sudden
change, which implies the breakdown of the adiabatic theorem and the Berry
phases in an adiabatic cyclic evolution are ill-defined near the anti-crossing point
in the spectrum.
\end{abstract}
\end{frontmatter}
\section{Introduction}

The Pancharatnam-Berry phase \cite{Pancharatnam,Berry}, or most commonly known
as Berry phase, has provided us a
very deep insight on the geometric structure of quantum mechanics. It is
worthwhile to note that Berry phase has attracted considerable interests in
quantum theory on account of giving rise to interesting observable
physical phenomena, implementing the operation of a universal quantum
logic gate in quantum computing \cite{Nazir,Kim,Leibfried,Jones}.
The most significant characteristic for the concept of Berry phase is the existence
of a continuous parameter space in which the state of the system varies
slowly along a closed cycle \cite{Berry, Holstein, XiaoRMP}. In particular, various
extensions of the phase have been considered \cite{Yi}, such as geometric phases for
mixed states \cite{Sjqvist}, for open systems \cite{Carollo}, and with a quantized driving field
\cite{Guridi, Bu, Chen, Larson, Larson1, Mart}. Furthermore, the concept of a geometric
phase has been generalized to the case of noneigenstates, which is
applicable to both linear and nonlinear quantum systems \cite{Wu}. In the
linear case, the geometric phase reduces to a statistical average of Berry
phases for the eigenstates, weighted by the probabilities that the system
finds itself in the eigenstates.

\begin{figure}[t]
\centering
\includegraphics[width=0.8\textwidth]{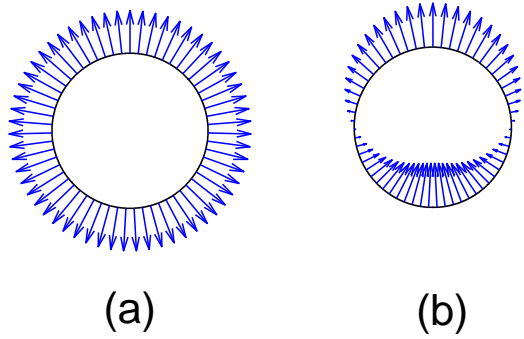}
\caption{(Color Online) The cross section of the geometric curvature
distributed on the surface of a unit sphere. (a) is the Berry curvature for
spin-$1/2$ particle in a magnetic field, the eigenstate $\Psi_1^+$ of JC
model, and the eigenstate $\Psi_1^2$ of the two-qubit Rabi model. (b) is the
vacuum-induced geometric curvature for the initial state $\left\vert
1\right\rangle \left\vert 0\right\rangle $ of JC model and $\left\vert
10\right\rangle \left\vert 0\right\rangle $ of the two-qubit Rabi model.}
\label{fig1}
\end{figure}

The Jaynes-Cummings (JC) model \cite{Jaynes}, or the quantum Rabi model \cite%
{Rabi} with the rotating wave approximation (RWA) that describes a spin-1/2
particle interacting with a single mode quantized field, plays an important role
in the cavity quantum electrodynamics. So Fuentes-Guridi et al. \cite{Guridi}
calculated the Berry phase of the JC model considering the quantum nature of
the field. A recent work \cite{Liu1} made a comparison study between the Berry
phases of the JC model with
and without the RWA where the Berry phase for the ground state varied from
zero under the RWA to nonzero values beyond the RWA. In addition, by means of
two unitary transformations, the Berry phase of the JC model without
the RWA, is presented in a simple and straightforward analytical method \cite%
{Deng}. Clearly, the Berry phases mentioned above are calculated using the
gauge dependent Berry connection. However, since the final result was gauge
independent, there should be a gauge independent way to calculate it. This
is provided by the Berry curvature that plays the role of a magnetic field
in the parameter space and is a more fundamental quantity than the Berry
connection. The example of spin-1/2 particle in a magnetic field
is often used to demonstrates the basic concepts and
important properties of the Berry phase \cite{Berry,Holstein,XiaoRMP}.
The corresponding curvature in its vector form is an effective
magnetic field in parameter space generated by a Dirac magnetic monopole of
strength $-1/2$ in the origin as shown in Fig. \ref{fig1}(a), where we present the
cross section for the field distribution on the surface of a unit sphere. Obviously,
it is homogeneous and isotropic in parameter space. Berry phase can then
be interpreted as the magnetic flux through the area whose boundary is the
closed loop. It only depends on the global property of the adiabatic evolutions,
which offers potential advantages against local fluctuations for implementing
geometric quantum gates \cite{Sjqvist1}. Motivated by this application of Berry phase
in phase-shift gate operation in quantum computation \cite{Nazir,Kim,Leibfried,Jones},
we investigate the effect of the geometric phase of the Rabi model from the viewpoint
of geometric connection and curvature.

In this paper we calculate the geometric phase of the two-qubit Rabi model where
two qubits interact with the quantized field inhomogeneously. In Sec II we first review the analytical
results for eigen solutions of the two-qubit Rabi model with the RWA according to
the conservation of total number of excitation and present the analytical expressions
of the Berry phase for the eigenstates. The procedure is as follows: From the connection
we derive the curvature, the integration of which gives the phase. Then we turn to the
geometric curvature and phase for noneigenstates in order to tackle the pure geometric
effect induced by the vacuum photon state. The system returns back to its initial state after
a period, which realizes a vacuum-to-vacuum cyclic evolution. We analyze
the corresponding geometric curvature field, in comparison with the example of
spin-$1/2$ particle. In Sec. III we study the Berry phase of the two-qubit Rabi model
beyond the RWA. The adiabatic approximation is adopted to derive the analytical results
for Berry phases of the eigenstates. For strong coupling case we truncate the Hilbert space
and calculate the phases numerically. The geometric phases for the evolution of noneigenstates
are compared with and without the RWA. Sec. IV summarizes our main findings.

\section{Geometric phase under the RWA}

\subsection{Berry phase for eigenstates}

The two-qubit Rabi model is described by the Hamiltonian ($\hbar =1$) \cite%
{Rabi}%
\begin{equation}
H=\omega _{c}a^{\dagger }a+\sum\limits_{j=1,2}\left( \frac{\omega _{j}}{2}%
\sigma _{j}^{z}+g_{j}\left( a^{\dagger }+a\right) \sigma _{j}^{x}\right) .
\label{HRWA}
\end{equation}%
Here $a^{\dagger }$ ($a$) is the bosonic creation (annihilation) operator of
field mode with frequency $\omega _{c}$, and $\omega _{j}$ denotes the
energy splitting of qubit $j$ described by Pauli matrices $\sigma _{j}^{x},
\sigma _{j}^{y}$ and $\sigma _{j}^{z}$. We notice that the dipole-field
coupling strength $g_{j}$ are not necessarily the same for the two qubits,
rendering a typical inhomogeneously coupled system.

One convenient way to handle the Hamiltonian is the RWA where the
counter-rotating terms are neglected in the weak coupling case. We begin
with the simplest case with $\omega _{2}=0$ and $g_{2}=0$, in which case the
Hamiltonian (\ref{HRWA}) is simplified to the JC model with eigenvalues
\begin{eqnarray}
E_{0} =-\omega _{1}/2,\ E_{k}^{\pm } =\omega _{c}\left( k-1/2\right) \pm \Omega _{k}/2,
\label{JCenergy}
\end{eqnarray}%
where $k$ denotes the total excitations of the qubit and the photons, the
Rabi frequencies $\Omega _{k}=\sqrt{\Delta^{2}+4g_{1}^{2}k}$, and the
detuning $\Delta =\omega _{1}-\omega _{c}$. Its eigenstates are
\begin{eqnarray}
\left\vert \Psi _{0}\right\rangle &=&\left\vert 0\right\rangle \left\vert
0\right\rangle ,  \nonumber \\
\left\vert \Psi _{k}^{+}\right\rangle &=&\cos \left( \theta _{k}/2\right)
\left\vert 1\right\rangle \left\vert k-1\right\rangle +\sin \left( \theta
_{k}/2\right) \left\vert 0\right\rangle \left\vert k\right\rangle ,  \nonumber
\\
\left\vert \Psi _{k}^{-}\right\rangle &=&\sin \left( \theta _{k}/2\right)
\left\vert 1\right\rangle \left\vert k-1\right\rangle -\cos \left( \theta
_{k}/2\right) \left\vert 0\right\rangle \left\vert k\right\rangle ,
\label{JCstate}
\end{eqnarray}%
where $\theta _{k}=\arccos \left( \Delta /\Omega_k\right)$, $\left\vert
1\right\rangle $ and $\left\vert 0\right\rangle $ are respectively the upper
and lower eigenstates of $\sigma _{1}^{z}$, and $\left\vert k\right\rangle $
is photon number state.

The Berry phase is known as a geometric effect when the wave function of the
system undergoes adiabatic evolution along a closed curve in the parameter
space \cite{Berry}. The phase change in the coupled state of qubits and
field is generated by introducing the phase shift operator $R(\phi
)=e^{-i\phi a^{\dagger }a}$, which is applied adiabatically to the
Hamiltonian of the system, i.e. $H\left( \phi \right) =R^{\dag }(\phi
)HR(\phi )$. In the simplest case, where the detuning and coupling strength
are fixed and the phase $\phi $ is varied slowly from $0$ to $2\pi $, each
eigenstate of the system undergoes a closed loop $\mathcal{C}$ on the Bloch
sphere and the Berry phase induced in this way is calculated as%
\begin{equation}
\gamma =i\int_{\mathcal{C}}d\phi \left\langle \Psi \right\vert R^{\dagger
}(\phi )\frac{d}{d\phi }R(\phi )\left\vert \Psi \right\rangle .
\label{berry}
\end{equation}%
For the eigenstates (\ref{JCstate}) we easily obtain
\begin{eqnarray}
\gamma _{0} =0, \
\gamma _{k}^{+} =\pi \left( 1-\cos \theta _{k}\right) +2\pi \left(
k-1\right) ,  \
\gamma _{k}^{-}=-\pi \left( 1-\cos \theta _{k}\right) +2\pi k.
\label{JCberry}
\end{eqnarray}

For the two-qubit Rabi model (\ref{HRWA}), we use the eigenstates
of the free qubits and free field as basis vectors and treat the Hamiltonian in the
Hilbert space
\begin{equation}
\chi =\left\{ \left\vert 11\right\rangle \left\vert n\right\rangle
,\left\vert 10\right\rangle \left\vert n\right\rangle ,\left\vert
01\right\rangle \left\vert n\right\rangle ,\left\vert 00\right\rangle
\left\vert n\right\rangle \right\}.  \label{space}
\end{equation}%
The product states of two qubits are defined as $\left\vert 11\right\rangle
=\left\vert 1\right\rangle _{1}\otimes \left\vert 1\right\rangle _{2}$, $%
\left\vert 10\right\rangle =\left\vert 1\right\rangle _{1}\otimes \left\vert
0\right\rangle _{2}$, $\left\vert 01\right\rangle =\left\vert 0\right\rangle
_{1}\otimes \left\vert 1\right\rangle _{2}$, $\left\vert 00\right\rangle
=\left\vert 0\right\rangle _{1}\otimes \left\vert 0\right\rangle _{2}$. The
space is shared with the operator $C=a^{\dagger }a+\left( \sigma
_{1}^{z}+\sigma _{2}^{z}+2\right) /2$, which again counts the total
excitations of the qubits and the photons. Denote the eigenvalues of $C$ as $%
k $, which can be used as quantum number to classify the energy eigenstates.
This leads immediately to a decomposition of the system Hilbert space into
subspaces, i.e. $\chi =\sum_{k=0}^{\infty }\oplus \chi _{k}$ with
\begin{eqnarray}  \label{subspace}
\chi _{0}&=&\left\{ \left\vert 00\right\rangle \left\vert 0\right\rangle
\right\} ,  \
\chi _{1}=\left\{ \left\vert 10\right\rangle \left\vert 0\right\rangle
,\left\vert 01\right\rangle \left\vert 0\right\rangle ,\left\vert
00\right\rangle \left\vert 1\right\rangle \right\} ,  \nonumber \\
\chi _{k}&=&\left\{ \left\vert 11\right\rangle \left\vert k-2\right\rangle
,\left\vert 10\right\rangle \left\vert k-1\right\rangle ,\left\vert
01\right\rangle \left\vert k-1\right\rangle ,\left\vert 00\right\rangle
\left\vert k\right\rangle \right\}.
\end{eqnarray}%
We note that except the state $\chi _{0}$, the subspace is three-fold
degenerate for $k=1$, and four-fold degenerate for $k > 1$. The fact that $C$
commutes $H$ makes it true that the subspace $k$ is invariant under $H$, and
the matrix representation of $H$ takes a block diagonal form $%
H=diag\{H_{0},H_{1},\cdots H_{k},\cdots \}$ with
\begin{eqnarray}
H_{0} &=&%
\begin{array}{c}
-\frac{\omega _{1}+\omega _{2}}{2}%
\end{array}%
,  \nonumber \\
H_{1} &=&\left(
\begin{array}{ccc}
\frac{\omega _{1}-\omega _{2}}{2} & 0 & g_{1} \\
0 & \frac{-\omega _{1}+\omega _{2}}{2} & g_{2} \\
g_{1} & g_{2} & \omega _{c}-\frac{\omega _{1}+\omega _{2}}{2}%
\end{array}%
\right) ,  \label{matrix1}
\end{eqnarray}%
and%
\begin{eqnarray}
H_{k} &=&\left(
\begin{array}{cccc}
-\omega _{c}+\frac{\omega _{1}+\omega _{2}}{2} & g_{2}\sqrt{k-1} & g_{1}%
\sqrt{k-1} & 0 \\
g_{2}\sqrt{k-1} & \frac{\omega _{1}-\omega _{2}}{2} & 0 & g_{1}\sqrt{k} \\
g_{1}\sqrt{k-1} & 0 & \frac{-\omega _{1}+\omega _{2}}{2} & g_{2}\sqrt{k} \\
0 & g_{1}\sqrt{k} & g_{2}\sqrt{k} & \omega _{c}-\frac{\omega _{1}+\omega _{2}%
}{2}%
\end{array}%
\right)  \nonumber \\
&&+\left( k-1\right) \omega _{c}I  \label{matrix2}
\end{eqnarray}%
with $I$ the identity matrix. The eigenstates can be solved by diagonalizing
the matrices $H_{k}$ in associated subspaces as%
\begin{eqnarray}
\left\vert \Psi _{0}\right\rangle &=&\left\vert 00\right\rangle \left\vert
0\right\rangle ,  \nonumber \\
\left\vert \Psi _{1}^{l}\right\rangle &=&b_{1}^{l}\left\vert 10\right\rangle
\left\vert 0\right\rangle +c_{1}^{l}\left\vert 01\right\rangle \left\vert
0\right\rangle +d_{1}^{l}\left\vert 00\right\rangle \left\vert
1\right\rangle ,  \nonumber \\
\left\vert \Psi _{k}^{l}\right\rangle &=&a_{k}^{l}\left\vert 11\right\rangle
\left\vert k-2\right\rangle +b_{k}^{l}\left\vert 10\right\rangle \left\vert
k-1\right\rangle+c_{k}^{l}\left\vert 01\right\rangle \left\vert k-1\right\rangle
+d_{k}^{l}\left\vert 00\right\rangle \left\vert k\right\rangle ,
\label{RWAfunction}
\end{eqnarray}%
where $l$ labels the different eigenstates with any given $k$ ($l=1,2,3$
with $k=1$ and $l=1,2,3,4$ with $k>1$). Substituting the expression of $%
\left\vert \Psi _{k}^{l}\right\rangle $ into Eq. (\ref{berry}), we obtain
the Berry phases for the eigenstates of two-qubit Rabi model
\begin{eqnarray}
\gamma _{0} =0,\ \gamma _{1}^{l} =\pi \left( 1-\cos \theta _{1}^{l}\right), \ \gamma _{k}^{l}=\mathop{\rm sgn}(s_{k})\pi \left( 1-\cos \theta
_{k}\right) +2\pi \left( k-1\right) ,  \label{berry1}
\end{eqnarray}%
where $\cos \theta _{1}^{l}=1-2\left\vert d_{1}^{l}\right\vert ^{2}$, $\cos
\theta _{k}^{l}=1-2\left\vert s_{k}^{l}\right\vert $ with $s_{k}=$ $%
\left\vert d_{k}^{l}\right\vert ^{2}-\left\vert a_{k}^{l}\right\vert ^{2}$
and $\mathop{\rm sgn}(x)$ is the sign function.

The coefficients of the eigenstates are tedious algebraic functions of
system parameters $\omega _{1,2}$ and $g_{1,2}$. As an example, in the case
of $\omega _{1}=\omega _{2}$, the eigenvalues for $k=1$ can be expressed as%
\begin{equation}
E_{1}^{l}=0,\left( -\Delta \pm \Theta _{1}\right) /2,  \label{Tenergy}
\end{equation}%
with $\Theta _{1}=\sqrt{\Delta ^{2}+4\left( g_{1}^{2}+g_{2}^{2}\right) }$.
We find $\theta _{1}^{1}=0$, $\theta _{1}^{2}=\arccos \left( \Delta /\Theta
_{1}\right) ,$ $\theta _{1}^{3}=\theta _{1}^{2}$ $+\pi $ and the three
eigenstates with $k=1$ can be simplified as
\begin{eqnarray}
\left\vert \Psi _{1}^{1}\right\rangle &=&\left\vert \varphi
_{0}^{-}\right\rangle , \
\left\vert \Psi _{1}^{2}\right\rangle =\cos \left( \theta
_{1}^{2}/2\right) \left\vert \varphi _{0}^{+}\right\rangle +\sin \left(
\theta _{1}^{2}/2\right) \left\vert \varphi _{1}\right\rangle ,  \nonumber \\
\left\vert \Psi _{1}^{3}\right\rangle &=&\sin \left( \theta
_{1}^{2}/2\right) \left\vert \varphi _{0}^{+}\right\rangle -\cos \left(
\theta _{1}^{2}/2\right) \left\vert \varphi _{1}\right\rangle ,
\label{TCstate}
\end{eqnarray}%
where $\left\vert \varphi _{0}^{\pm }\right\rangle $ and $\left\vert \varphi
_{1}\right\rangle $ are the eigenstates of the uncoupled system and
expressed as%
\begin{eqnarray}
\left\vert \varphi _{0}^{+ }\right\rangle &=&\left( \cos \alpha \left\vert
10\right\rangle + \sin \alpha \left\vert 01\right\rangle \right) \left\vert
0\right\rangle , \nonumber \\
\left\vert \varphi _{0}^{- }\right\rangle &=&\left( \sin \alpha \left\vert
10\right\rangle - \cos \alpha \left\vert 01\right\rangle \right) \left\vert
0\right\rangle ,  \
\left\vert \varphi _{1}\right\rangle =\left\vert 00\right\rangle
\left\vert 1\right\rangle .  \label{spin}
\end{eqnarray}%
with $\cos \alpha =g_{1}/\sqrt{g_{1}^{2}+g_{2}^{2} }$. Furthermore, the
corresponding Berry phases are respectively
\begin{eqnarray}
\gamma _{1}^{1}=0, \ \gamma _{1}^{2}=\pi\left(1-\cos \theta _{1}^{2}\right),
\ \gamma _{1}^{3}=\pi \left( 1+\cos \theta _{1}^{2}\right).  \label{Berry3}
\end{eqnarray}

Here we discuss some features of the Berry phases for some exceptional
eigenstates of the system. First, similar to the case of JC model, the
vacuum state $\left\vert \Psi _{0}\right\rangle =\left\vert 00\right\rangle
\left\vert 0\right\rangle $ is the real ground state with energy $%
E_{0}=-\left( \omega _{1}+\omega _{2}\right) /2$ and the corresponding Berry
phase is trivially zero. Second, in the case of $\omega _{1}=\omega _{2}$
and $k=1$, it is easy to show that the eigenstate $\left\vert \psi
\right\rangle =\left( g_{2}\left\vert 10\right\rangle -g_{1}\left\vert
01\right\rangle \right) \left\vert 0\right\rangle /\sqrt{g_{1}^{2}+g_{2}^{2}}
$ acquires no geometric phase, too. These two eigenstates share a common
feature that only vacuum state of the bose field $\left\vert 0\right\rangle $
is involved. Finally, for two identical qubits $\omega _{1}=\omega _{2}$ and
$g_{1}=g_{2}$, the spin singlet states $\left\vert \psi _{n}\right\rangle =1/%
\sqrt{2}\left( \left\vert 10\right\rangle -\left\vert 01\right\rangle
\right) \left\vert n\right\rangle $ (for any $n$) are eigenstates of the
system. The corresponding Berry phases are integer multiples of $2\pi $,
which would not affect the wave functions. The resultant phases for the
eigenstates other than the above-mentioned three exceptional cases are
generally non-zero, which are attributed to the interaction between the
qubits and the bosonic field.

\subsection{Berry curvature for eigenstates}

The eigenstates Eq. (\ref{JCstate}) are similar to those of a spin-$\frac{1}{%
2}$ particle in an external magnetic field, which describes a number of
physical systems in condensed matter physics \cite{XiaoRMP}. To better
understand the geometric properties of the parameter space, we describe the
Berry phase of the JC model in an alternative way. First we define the Berry
connection or the Berry vector potential as
\begin{equation}
\mathbf{A}_{n}=i\left\langle \Phi _{n}\right\vert \nabla _{\mathbf{\lambda }%
}\left\vert \Phi _{n}\right\rangle  \label{econnection}
\end{equation}%
where $\Phi _{n}$ is the eigenstate of the system that depends on time
through a set of parameters denoted by $\mathbf{\lambda =}\left\{ \lambda
_{1},\lambda _{2},\cdots \right\} $. For the JC model, the parameter space
is spanned by $\theta =\theta _{1}$ and $\phi $ and the Berry connection is
calculated for the eigenstates $\Psi _{1}^{\pm }$ as $A_{\theta }^{\pm }=0$
and $A_{\phi }^{+}=\sin ^{2}\left( \theta _{1}/2\right) $, $A_{\phi
}^{-}=\cos ^{2}\left( \theta _{1}/2\right) $, which is obviously gauge
dependent. It is thus useful to define a gauge-field tensor, known as the
Berry curvature which can be derived from the Berry vector potential
\begin{equation}
F_{\theta \phi }=\partial _{\theta }A_{\phi }-\partial _{\phi }A_{\theta }.
\label{curvature}
\end{equation}%
It provides a local description of the geometric properties of the parameter
space. The Berry curvature is gauge invariant and thus observable. It is a
more fundamental quantity than the Berry phase, just as has been shown in
Refs. \cite{Kura, Shi} that the Berry curvature directly
participates in the dynamics of the adiabatic parameters and the orbital
magnetization contains a Berry curvature contribution of topological origin.
For the two eigenstates $\Psi _{1}^{\pm }$ we have $F_{\theta \phi }^{\pm
}=\pm \frac{1}{2}\sin \theta _{1}$. According to Stokes' theorem the Berry
phase is the integral of the curvature on the surface $\mathcal{S}$ enclosed
by the loop $\mathcal{C}$ on the Bloch sphere, i.e.
\begin{equation}
\gamma =\int\nolimits_{\mathcal{S}}F_{\theta \phi }d\theta d\phi ,
\label{eberry}
\end{equation}%
which reproduces the result Eq. (\ref{JCberry}).

One may on the other hand parametrize the parameter space in the spherical
coordinates $r, \theta $ and $\phi $, in which case the Berry curvature can
be recast into a vector form
\begin{equation}
\vec{F}_{n}=\nabla \times \mathbf{A}_{n},
\end{equation}%
which can be viewed as an effective magnetic field in the parameter space.
The result for the eigenstate $\Psi _{1}^{+}$ is
\begin{equation}
\vec{F}^{+}=\frac{1}{2r^{2}}\vec{e}_{r},  \label{F}
\end{equation}%
where the polar angle $\theta _{1}$ varies from $0$ to $\pi /2$ for $\Delta
>0$ and from $\pi /2$ to $\pi $ for $\Delta <0$, and the radius $r=\Omega
_{1}/2$ changes with the coupling strength $g_1$. When the field is mapped
onto the unit sphere, the curvature for the JC model is isotropic just as in
the example of spin-$1/2$ particle, which is recognized as a magnetic field
generated by a monopole at the origin $r=0$ as shown in Fig. 1(a) \cite%
{XiaoRMP}. The direction of the Berry curvature field for the eigenstate $%
\Psi_{1}^{-}$ is opposite to the field $\vec{F}^{+}$.

The Berry connection and the Berry curvature for the two-qubit Rabi model
can be calculated in the same way by mapping the parameter space into the
model of spin-$1/2$ particle. For the three eigenstates (\ref{TCstate}) with
$k=1$ the Berry connection as a function of $\theta =\theta _{1}^{2} $ and $%
\phi $ is $A_{\theta }^{l}=0$ and $A_{\phi }^{1}=0, A_{\phi }^{2}=\sin
^{2}\left( \theta _{1}^{2}/2\right), A_{\phi }^{3}=\cos ^{2}\left( \theta
_{1}^{2}/2\right) $. The Berry curvature can be derived easily as $F_{\theta
\phi }^{1}=0,F_{\theta \phi }^{2}=\frac{1}{2}\sin \theta _{1}^{2}$, and $%
F_{\theta \phi }^{3}=-\frac{1}{2}\sin \theta _{1}^{2}$, from which one can
easily recover the Berry phase in Eq. (\ref{Berry3}). In spherical
coordinates the Berry curvature for the eigenstate $\Psi _{1}^{2}$ takes
the same form as Eq. (\ref{F}) with, however, the radius replaced by $%
r=\Theta _{1}/2$. The Berry curvatures for the eigenstates
$\Psi _{1}^{2}$ and $\Psi _{1}^{3}$ thus show exactly the same structures as
those for the eigenstates $\Psi _{1}^{+}$ and $\Psi _{1}^{-}$ of the JC
model, respectively. However, the Berry curvature for the eigenstate $%
\Psi_{1}^{1}$ is constantly zero in the whole unit sphere because only
vacuum state of the bose field $\left\vert 0\right\rangle $ is involved.

\subsection{Geometric curvature and phase for noneigenstates}

We now turn to the geometric curvature and phase for noneigenstates of the
JC model in order to tackle the pure geometric effect induced by the vacuum
photon state. Staring from the initial state $\left\vert 1\right\rangle
\left\vert 0\right\rangle $ with a qubit in the excited state $\left\vert
1\right\rangle $ and field in the vacuum state $\left\vert 0\right\rangle $,
which happens to be a superposition of two eigenstates $\Psi _{1}^{\pm }$ of
the JC model
\begin{equation}
\left\vert 1\right\rangle \left\vert 0\right\rangle =\cos \left( \theta
_{1}/2\right) \left\vert \Psi _{1}^{+}\right\rangle
+\sin \left( \theta
_{1}/2\right) \left\vert \Psi _{1}^{-}\right\rangle,
\end{equation}
the time-dependent state after an adiabatic and cyclic evolution is shown to
be \cite{Minghao}
\begin{eqnarray}
\left\vert \Psi \left( T\right) \right\rangle &=&\cos \left( \theta
_{1}/2\right) e^{-i\int_{0}^{T}E_{1,+}^{\prime }dt}e^{i\gamma
_{1}^{+}}\left\vert \Phi _{1}^{+}\right\rangle \nonumber \\
&+&\sin \left( \theta _{1}/2\right) e^{-i\int_{0}^{T}E_{1,-}^{\prime
}dt}e^{i\gamma _{1}^{-}}\left\vert \Phi _{1}^{-}\right\rangle ,
\label{generalstate1}
\end{eqnarray}%
where $E_{1,\pm }^{\prime }=E_{1}^{\pm }-i\left\langle \Psi _{1}^{\pm
}\right\vert \frac{dR^{\dag }}{dt}R\left\vert \Psi _{1}^{\pm }\right\rangle $
and $\left\vert \Phi _{1}^{\pm }\right\rangle =R\left\vert \Psi _{1}^{\pm
}\right\rangle $ with $R\left( T\right) =R\left( 0\right) $. The evolution
of the noneigenstate exhibits clear recurrence behavior with a period $%
T=2\pi /\Omega _{1}$, that is, the state $\left\vert \Psi \left( t\right)
\right\rangle $ returns back to the initial state at subsequent time $qT$ ($%
q=1,2, \cdots$) with an additional phase factor, $e^{i\Gamma }\left\vert
1\right\rangle \left\vert 0\right\rangle $. This is a beautiful realization
of the vacuum-to-vacuum evolution. Remove the dynamic part from the total
phase $\Gamma $, the Aharonov-Anandan phase (AA) phase of the evolution is
obtained as \cite{Minghao}
\begin{equation}
\beta =\cos ^{2}\left( \theta _{1}/2\right) \left( \gamma _{1}^{+}+2q\pi
\right) +\sin ^{2}\left( \theta _{1}/2\right) \gamma _{1}^{-}.
\label{AAphase}
\end{equation}%
It was found that the average photon number of the state $\left\vert \Psi
\left( t\right) \right\rangle $ in a period is related to the Berry phases $%
\gamma _{1}^{\pm }$ by
\begin{eqnarray}
P &=&\frac{1}{T}\int_{0}^{T}\left\langle \Psi \left( t\right) \right\vert
a^{\dag }a\left\vert \Psi \left( t\right) \right\rangle dt  \nonumber \\
&=&\left( \cos ^{2}\left( \theta _{1}/2\right) \gamma _{1}^{+}+\sin
^{2}\left( \theta _{1}/2\right) \gamma _{1}^{-}\right) /2\pi .
\label{photon}
\end{eqnarray}

Here we discuss the geometric phase for the adiabatic and cyclic evolution
of noneigenstates $\left\vert 1\right\rangle \left\vert 0\right\rangle $
according to a formal definition given by Wu \textit{et al.} \cite{Wu}. For
the linear systems, it is defined as a statistical average of Berry phases
for the eigenstates, weighted by the probabilities $\left\vert
a_{n}\right\vert ^{2}$ that the system finds itself in the eigenstates
\begin{equation}
\gamma =\sum_{n}\left\vert a_{n}\right\vert ^{2}\oint d\mathbf{\lambda }%
\cdot \mathbf{A}_{n}.  \label{gberry}
\end{equation}%
Interestingly, this kind of weighted summation of Berry phases has already
been applied in real physical systems \cite{Thouless, Yao}. Our scheme is
based on Eq. (\ref{gberry}) and we define the geometric connection for
noneigenstates as
\begin{equation}
\mathbf{A=}\sum_{n}\left\vert a_{n}\right\vert ^{2}\mathbf{A}_{n}.
\label{cn}
\end{equation}%
We find for noneigenstate $\left\vert 1\right\rangle \left\vert
0\right\rangle $, this connection is $A_{\theta }=0,A_{\phi }=\frac{1}{2}%
\sin ^{2}\theta _{1}$ and the corresponding geometric curvature is
calculated from the definition (\ref{curvature}) as
\begin{equation}
F_{\theta \varphi }=\frac{1}{2}\sin 2\theta _{1}.  \label{jcgp}
\end{equation}
Based on the fact that the adiabatic evolution of noneigenstate forms a
closed loop in the parameter space at time $T$, the integral of the
curvature Eq. (\ref{eberry}) on the surface $\mathcal{S}$ defines a new
geometric phase $\gamma$, the result for state $\left\vert 1\right\rangle
\left\vert 0\right\rangle $ being just the same as Eq. (\ref{gberry})
\begin{equation}
\gamma =\frac{1}{2}\pi \left( 1-\cos 2\theta _{1}\right) .  \label{JCGP}
\end{equation}%
We conclude that while the AA phase $\beta $ describes the pure geometric
phase acquired in the cyclic evolution, the geometric phase $\gamma $ is
related to the average photon number through
\begin{equation}
P=\gamma /2\pi .  \label{Relationship}
\end{equation}%
Obviously, the value of $\gamma $ can not exceed $\pi $, i.e., $P\leq 1/2,$
because the total excitations of the qubit and the photon is a conserved
quantity with $k=1$ in the process of time evolution. It is more intuitive
to consider the geometric curvature (\ref{jcgp}) in the spherical
coordinates
\begin{equation}
\vec{F}=\frac{\cos \theta _{1}}{r^{2}}\vec{e}_{r},  \label{sphere2}
\end{equation}%
which acquires an additional polar angle dependent factor $\cos\theta_1$.
The distribution of the geometric curvature for the noneigenstate $%
\left\vert 1\right\rangle \left\vert 0\right\rangle $ is shown in Fig. \ref%
{fig1}(b) and we find the curvature is rotationally symmetric about $z$%
-axis. Interestingly, while the curvature is still centrifugal in the
northern hemisphere, the factor $\cos \theta _{1}$ reverses the direction of
the field in the southern hemisphere and the corresponding geometric
curvature field is opposite to the direction of the radius $\vec{e}_{r}$.
This can be recognized as a statistical average of the magnetic field generated
simultaneously by two monopoles with opposite magnetic charges.
The maximum field strength lies in the north and south poles, which is
chosen as the reference and set to be unity, while on the equator the field
is zero.

For the two-qubit Rabi model, the initial state $\left\vert 10\right\rangle
\left\vert 0\right\rangle $ with the field in the vacuum state can be
expressed as a superposition of eigenstates of the system
\begin{equation}
\left\vert \Psi \left( 0\right) \right\rangle =\sum_{l}b_{1}^{l}\left\vert
\Psi _{1}^{l}\right\rangle.  \label{Gstate}
\end{equation}%
with
\begin{eqnarray}
b_1^1 =\sin \alpha, \ b_1^2 &=&\cos \alpha \cos \left( \theta _{1}^{2}/2\right), \
b_1^3 =\cos \alpha \sin \left( \theta _{1}^{2}/2\right).  \label{b1l}
\end{eqnarray}%
The time-dependent state after an adiabatic and cyclic evolution is similar
to the case of JC model \cite{Minghao}
\begin{equation}
\left\vert \Psi \left( T\right) \right\rangle
=\sum_{l}b_{1}^{l}e^{-i\int_{0}^{T}E_{1 l}^{\prime }\left( t^{\prime
}\right) dt}e^{i\gamma _{1}^{l}}\left\vert \Phi _{1}^{l}\right\rangle,
\label{rabistate}
\end{equation}%
where the instantaneous eiegnstates are $\left\vert \Phi
_{1}^{l}\right\rangle =R\left\vert \Psi _{1}^{l}\right\rangle $ and the
corresponding eiegnenergies are
\begin{equation}
E_{1l}^{\prime } =E_{1}^{l}-i\left\langle \Psi _{1}^{l}\right\vert \frac{%
dR^{\dag }}{dt}R\left\vert \Psi _{1}^{l}\right\rangle.
\end{equation}%
Consider for example the special case of $\omega _{1}=\omega _{2}$, the
eigenenergies and eigenstates of which are given by the Eqs. (\ref{Tenergy})
and (\ref{TCstate}) in the previous section. Interestingly, if the ratio of
the two eigenvalues is a rational number
\begin{equation}
E_{1}^{2}/E_{1}^{3}=p/q,  \label{condition}
\end{equation}%
where $p$ and $q$ are integers, at subsequent time $T=2p\pi /E_{1}^{2}=2q\pi
/E_{1}^{3}$ the noneigenstate returns back to the initial state with an
additional phase factor $\Gamma $, i.e. $\left\vert \Psi \left( T\right)
\right\rangle =e^{i\Gamma }\left\vert 10\right\rangle \left\vert
0\right\rangle $. In a recent work \cite{Ballester} Ballester \textit{et al.}
introduced an effective qubit-cavity system tuning the parameters of the
external drivings, where the frequency $\omega _{c}$ is in a magnitude of $%
2\pi \times 10$ MHz and in near resonance with the frequency of the qubit in
the strong driving limit. For example, for a detuning $\Delta =0.01\omega
_{c} $ and weak coupling $g_{1}=g_{2}=0.01\omega _{c}$, we have $p/q=-1/2$
and $T=5 \mu s$. This is the vacuum-to-vacuum evolution in the two-qubit
case and the geometric phase in the evolution is induced by pure vacuum
state of photons. The AA phase in the evolution is again obtained by
removing the dynamical phase from $\Gamma $
\begin{eqnarray}
\beta &=&\cos ^{2}\alpha \cos ^{2}\left( \theta _{1}^{2}/2\right) \left(
\gamma _{1}^{2}+2p\pi \right)+\cos ^{2}\alpha \sin ^{2}\left( \theta _{1}^{2}/2\right) \left( \gamma
_{1}^{3}+2q\pi \right) .  \label{twoAA}
\end{eqnarray}
The connection for the state $\left\vert 10\right\rangle \left\vert
0\right\rangle $ is $A_{\theta }=0,A_{\phi }=\frac{1}{2}\sin ^{2}\theta
_{1}^{2}\cos ^{2}\alpha $, from which the geometric curvature is obtained as
\begin{equation}
F_{\theta \phi }=\frac{1}{2}\sin 2\theta _{1}^{2}\cos ^{2}\alpha .
\label{trc}
\end{equation}%
The geometric phase is derived by a surface integral according to Eq. (\ref%
{eberry})
\begin{equation}
\gamma =\frac{1}{2}\pi \cos ^{2}\alpha \left( 1-\cos 2\theta _{1}^{2}\right),
\label{gp1}
\end{equation}%
and related to the AA phase $\beta $ through
\begin{equation}
\beta =\gamma +2\pi \cos ^{2}\alpha \left[ p\cos ^{2}\left( \theta
_{1}^{2}/2\right) +q\sin ^{2}\left( \theta _{1}^{2}/2\right) \right] .
\label{relationship}
\end{equation}%
The vacuum induced geometric phase $\gamma $ is again related to the average
photon number in a period $T$ through Eq. (\ref{Relationship}) and thus can
be measured by counting the photons during the cyclic evolution. The average
photon number $P$ in the two-qubit model is limited to a maximum value which
depends on the ratio of the coupling strength $\cos ^{2}\alpha $. For
homogeneous coupling $g_{1}=g_{2}$, we find $P\le 1/4$. The geometric
curvature in the spherical coordinates
\begin{equation}
\vec{F}=\frac{\cos ^{2}\alpha \cos \theta^2_{1}}{r^{2}}\vec{e}_{r}
\label{sphere4}
\end{equation}%
shows exactly the same distribution in the parameter space as the single
qubit case in Fig. 1(b) once we set the maximum field strength to be unity.

\begin{figure}[t]
\centering
\includegraphics[width=0.8\textwidth]{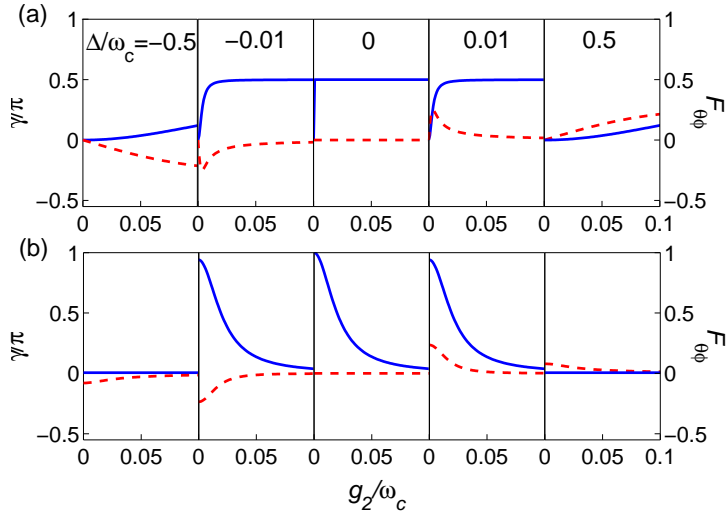}
\caption{(Color Online) The geometric curvature (red dashed curves) and
geometric phase (blue solid curves) for an adiabatic and cyclic evolution of
the initial state $\left\vert \protect\psi \left( t\right) \right\rangle $
as a function of the coupling strength $g_{2}/\protect\omega _{c}$ with (a) $%
g_{1}=g_{2}$ and (b) $g_{1}=0.02\protect\omega _{c}$ .}
\label{fig2}
\end{figure}

We illustrate in Fig. \ref{fig2} the geometric curvature $F_{\theta \phi }$
in (\ref{trc}) and the geometric phase $\gamma $ in (\ref{gp1}) as a
function of the coupling strength $g_2/\omega_c$ for an adiabatic and cyclic
evolution of the initial state $\left\vert 10\right\rangle \left\vert
0\right\rangle $. Several typical detuning cases are shown for homogeneous
and inhomogeneous coupling systems. While the geometric phases are always
positive definite, the respective geometric curvatures are of opposite sign
for red and blue detunings, as can be seen from Fig 1(b). For homogeneous
coupling case, while $\gamma $ increases gradually with the coupling
strength in the case of far off-resonant interaction, it reaches the maximum
value $\pi/2$ very quickly near the resonance. The geometric curvature, on
the other hand, develops a peak or a valley for small detuning at $%
g_2/\omega_c=\Delta/\sqrt{8}$, though it depends on $g$ monotonically for
large ones. For inhomogeneous coupling shown in Fig. \ref{fig2}(b), the
maximum geometric phase acquired in the evolution is $\pi$, which occurs in
the single qubit system for exact resonant condition. The phase and
curvature are suppressed to nearly zero for non-equal coupling strengths,
corresponding to the circle on the equator in Fig 1(b).

Experiments have shown that many engineering systems enable us to access
coupling strengths and detunings outside the regime in which the RWA is
valid, such as circuit QED experiments with superconducting qubits coupled
to LC and waveguide resonators \cite{Schuster, Forn-Diaz, Niemczyk,
Abdumalikov} and Cooper-pair boxes or Josephson phase qubits coupled to
nanomechanical resonators \cite{Bouchiat, Nakamura, Wallraff, Hofheinz,
LaHaye}. In next section we study the Berry phase beyond the RWA, which
shows very different feature from the RWA results.

\section{Berry phase beyond the RWA}

The Hamiltonian for the two-qubit Rabi model beyond the RWA fails to commute
with the total excitation operator $C$ and the block diagonal form of the
matrix $H$ breaks down. One would conclude that the appearance of the
counter-rotating terms makes it impossible to solve the Hamiltonian exactly.
However, the employment of Bargman-space technique, or displaced Fock state
provides a way to solve the model analytically \cite{Liu2, He, Peng, Mao}. The
adiabatic approximation proves to be an excellent way to treat the
eigenenergies when the transition frequency of the qubit $\omega _{j}$ is
much smaller than the frequency of bose field $\omega _{c}$ and the coupling
strength enters into the ultrastrong coupling regime. The eigenstates under
this approximation have the form%
\begin{equation}
\left\vert \psi _{n}^{\kappa \pm }\right\rangle =\frac{1}{\sqrt{2}}\left(
d_{1n}^{\kappa \pm },d_{2n}^{\kappa \pm },(-1)^{n}\kappa d_{2n}^{\kappa \pm
},(-1)^{n}\kappa d_{1n}^{\kappa \pm }\right) ^{T},  \label{SB}
\end{equation}%
which are expanded in the displaced Fock state basis $\left( \left\vert
11\right\rangle \left\vert n\right\rangle _{A_{1}},\left\vert
10\right\rangle \left\vert n\right\rangle _{A_{2}},\left\vert
01\right\rangle \left\vert n\right\rangle _{A_{3}},\left\vert
00\right\rangle \left\vert n\right\rangle _{A_{4}}\right) $. Here $\kappa
=\pm 1$ labels the parity, the displaced Fock states are defined as $%
A_{i}^{\dag }A_{i}\left\vert n\right\rangle _{A_{i}}=n\left\vert
n\right\rangle _{A_{i}}$ with $A_{i}=a+\beta _{i}$, and $\beta _{1}=-\beta
_{4}=\left( g_{1}+g_{2}\right) /\omega _{c}$ and $\beta _{2}=-\beta
_{3}=\left( g_{1}-g_{2}\right) /\omega _{c}$. The coefficients in Eq. (\ref%
{SB}) are
\begin{equation*}
d_{1n}^{\kappa \pm }=\xi _{n}^{\kappa \pm }\sqrt{\frac{1}{1+\left( \xi
_{n}^{\kappa \pm }\right) ^{2}}},d_{2n}^{\kappa \pm }=-\sqrt{\frac{1}{%
1+\left( \xi _{n}^{\kappa \pm }\right) ^{2}}},
\end{equation*}%
with
\begin{eqnarray*}
\xi _{n}^{\kappa \pm } &=&\Omega _{n}^{\kappa }/\left( \left( \beta
_{2}^{2}-\beta _{1}^{2}\right) \omega _{c}/2\mp \mu _{n}^{\kappa }\right)  \\
\mu _{n}^{\kappa } &=&\sqrt{\left( \Omega _{n}^{\kappa }\right) ^{2}+\omega
_{c}^{2}\left( \beta _{1}^{2}-\beta _{2}^{2}\right) ^{2}/4}, \\
\Omega _{n}^{\kappa } &=&-\frac{\omega _{1}}{2}\left[ _{A_{1}}\left\langle
n|n\right\rangle _{A_{2}}\right] -\kappa \left( -1\right) ^{n}\frac{\omega
_{2}}{2}\left[ _{A_{1}}\left\langle n|n\right\rangle _{A_{3}}\right] .
\end{eqnarray*}%
We find the corresponding eigenergies for the analytical eigenstates (\ref%
{SB}) are expressed as
\begin{equation}
E_{n}^{\kappa \pm }=n\omega _{c}-\left( \beta _{1}^{2}+\beta _{2}^{2}\right)
\omega _{c}/2\pm \mu _{n}^{\kappa },  \label{zero app}
\end{equation}%
and the corresponding Berry phases are calculated as
\begin{equation}
\gamma _{n}^{\kappa \pm }=\pi \left( 1-\cos \theta _{n}^{\kappa \pm }\right)
+2n\pi ,  \label{Bberry3}
\end{equation}%
with $\theta _{n}^{\kappa \pm }=2\arcsin \sqrt{\beta _{1}^{2}\left(
d_{1n}^{\kappa \pm }\right) ^{2}+\beta _{2}^{2}\left( d_{2n}^{\kappa \pm
}\right) ^{2}}$, which are accurate in the weak coupling case with $g_{1,2}$
up to $0.02\omega _{c}$. For even stronger coupling we resort to numerical
scheme described as following. The eigenstates are expanded in the truncated
displaced Fock space as
\begin{eqnarray}
\left\vert \Psi ^{\kappa }\right\rangle  &=&\sum_{n=0}^{M}\left[
d_{1n}^{\kappa }\left\vert 11\right\rangle \left\vert n\right\rangle
_{A_{1}}+d_{2n}^{\kappa }\left\vert 10\right\rangle \left\vert
n\right\rangle _{A_{2}}\right.   \nonumber \\
&+&\left. \kappa \left( -1\right) ^{n}\left( d_{2n}^{\kappa
}\left\vert 01\right\rangle \left\vert n\right\rangle
_{A_{3}}+d_{1n}^{\kappa }\left\vert 00\right\rangle \left\vert
n\right\rangle _{A_{4}}\right) \right] ,  \label{Beigenstate}
\end{eqnarray}%
and the Berry phase for eigenstates $\left\vert \Psi^\kappa \right\rangle $ is by
definition (\ref{berry}) calculated as
\begin{eqnarray}
\gamma ^{\kappa } &=&2\pi \sum\limits_{n=0}^{M}\left[ \left( n+\beta
_{1}^{2}\right) (d_{1n}^{\kappa })^{2}+\left( n+\beta _{2}^{2}\right)
(d_{2n}^{\kappa })^{2})\right.   \nonumber \\
&-&\left. 2\sqrt{n+1}\left( \beta _{1}d_{1,n+1}^{\kappa }d_{1n}^{\kappa
}+\beta _{2}d_{2,n+1}^{\kappa }d_{2n}^{\kappa }\right) \right] .
\label{Bberry}
\end{eqnarray}%
Here the coefficients $d$'s are obtained numerically by setting the
truncation number as $M=50$ such that the calculation is done in a closed
subspace $\left\vert n\right\rangle _{A_{i}}$ with $n=0,1,2\cdots ,M$.

Before we present the numerical results for the Berry phase of the
eigenstates, we first give the result for the exceptional spectrum of the
two-qubit model. First, the spin singlet state $\psi_n$ for two identical
qubits remains eigenstate of the system beyond the RWA, which acquires a
geometric phase $2n\pi$ as in RWA. In addition, in the homogeneous coupling
case $g_{1}=g_{2}$, there exists a constant analytical solution $E=\omega
_{c}$ \cite{Peng,Mao} corresponding to either the even parity eigenstate
\begin{equation*}
\left\vert \psi _{e}\right\rangle =\left( q_{e}\left( \left\vert
10\right\rangle -\left\vert 01\right\rangle \right) |1\rangle +\left\vert
11\right\rangle |0\rangle \right) /\sqrt{2q_{e}^{2}+1},
\end{equation*}%
for the symmetric detuning with $\omega _{1}+\omega _{2}=2\omega _{c}$
(suppose $\omega _{1}>\omega _{2}$), or the odd parity eigenstate
\begin{equation*}
\left\vert \psi _{o}\right\rangle =\left( q_{o}\left( \left\vert
00\right\rangle -\left\vert 11\right\rangle \right) |1\rangle +\left\vert
10\right\rangle |0\rangle \right) /\sqrt{2q_{o}^{2}+1},
\end{equation*}%
for the asymmetric detuning with $\omega _{1}-\omega _{2}=2\omega _{c}$ and $%
q_{e,o}=2g_{1}/\left( \omega _{1}\mp \omega _{2}\right) $. These two states
contain at most one photon and we have an exact result for their Berry
phases as
\begin{equation}
\gamma_{e,o}=\pi \left( 1-\cos \theta_{e,o}\right),  \label{Bberry2}
\end{equation}%
where $\cos \theta_{e,o}=\left( 1-2q_{e,o}^{2}\right) /\left(
1+2q_{e,o}^{2}\right) $. Both the vacuum state and the one-photon state of
the quantized field are involved in $\gamma_{e,o}$, which is analogous to
the Berry phase for the eigenstates (\ref{JCstate}) and (\ref{TCstate}) for $%
k=1$.

\begin{figure}[t]
\centering
\includegraphics[width=0.5\textwidth]{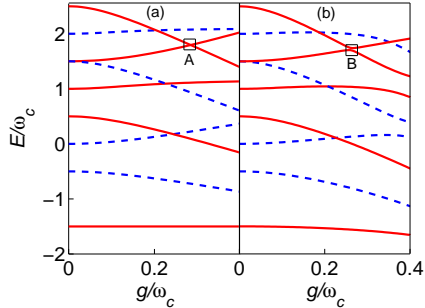}
\caption{(Color Online) Comparison of the energy under the RWA (a) with
beyond the RWA (b) depending on the dimensionless coupling $g/\protect\omega %
_{c}$ in the even (red solid lines) or odd (blue dashed lines) parity
subspaces, where we set $g_{1}=g_{2}=g$, $\protect\omega _{1}=\protect\omega %
_{2}$ and $\Delta =0.5\protect\omega _{c}.$}
\label{fig3}
\end{figure}
\begin{figure}[t]
\centering
\includegraphics[width=0.5\textwidth]{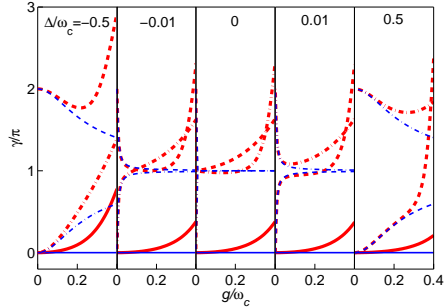}
\caption{(Color Online) Comparison of the Berry phases of the two-Rabi model
under RWA (thin blue curves) and beyond the RWA (bold red curves) as a
function of the coupling strength $g/\protect\omega _{c}$, where the solid,
dashed, and dashed-dotted curves correspond to the states $\Psi _{0},\Psi
_{1}^{2},\Psi _{1}^{3}$ in RWA and the corresponding numerical eigenstates
beyond RWA.}
\label{fig4}
\end{figure}

\begin{figure}[t]
\centering
\includegraphics[width=0.5\textwidth]{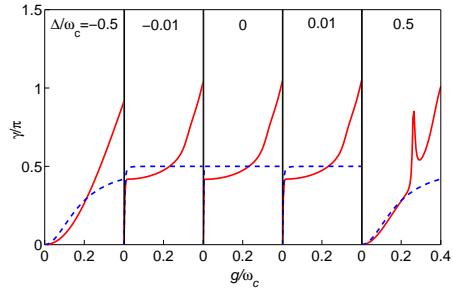}
\caption{(Color Online) Comparison of the geometric phases for
noneigenstates under the RWA (blue dashed curves) with beyond the RWA (red
solid curves) depending on the coupling strength $g/\protect\omega _{c}$.
The anomalous sudden change for $\Delta =0.5\protect\omega _{c}$ around $%
g\sim 0.265\protect\omega _{c}$ is attributed to the anti-crossing point (B)
in Fig. \protect\ref{fig3}(b).}
\label{fig5}
\end{figure}

We illustrate in Fig. \ref{fig3} the first few eigenergies of the
homogeneously coupling system with and without the RWA as a function of
coupling strength $g/\omega _{c}$ in the even and odd parity subspaces,
where the eigenenergies $E_{n}=n\omega _{c}$ corresponding to the spin
singlet states $\psi _{n}$ are not shown. Particularly we have marked a
crossing point (A) of energy levels for different values of $k$ in RWA and
an anti-crossing point (B) within the even parity subspace beyond RWA. It is
easily show that during the adiabatic evolution the quantum transition
between different subspace $k$ is forbidden in RWA, i.e. $\left\langle \Psi
_{k}|R^{\dag }H(\phi )R|\Psi _{k^{\prime }}\right\rangle =\delta
_{kk^{\prime }}$. The degeneracy at (A) thus would not affect the
calculation of the Berry phases. The nearly degeneracy at the anti-crossing
point (B) beyond RWA, on the other hand, will break the validity of the
adiabatic theorem as the transition between states in the same parity space
(in the case shown in Fig. 3(b), the even parity space) is not negligible,
and the condition for the adiabatic approximation $\left\vert \left\langle
\Psi _{n}^{\kappa }|R^{\dag }\dot{H}(\phi )R|\Psi _{m}^{\kappa
}\right\rangle /(E_{n}^{\kappa }-E_{m}^{\kappa })^{2}\right\vert \ll 1$ is
not applicable. We shall see below that these anti-crossing points lead to
nontrivial contribution in the geometric phase. It demonstrates that the RWA
indeed only works well in the weak coupling regime and the counterpart of
the real ground state energy $E_{0}$ is not any more a constant beyond RWA.
We choose to compare the Berry phases for three eigenstates $\Psi _{0},\Psi
_{1}^{2},\Psi _{1}^{3}$ in RWA and the corresponding numerical eigenstates
beyond RWA in Fig. \ref{fig4}, the eigenenergies of which are just the
lowest solid (red) and the two lowest dashed (blue) lines for the detuning $%
\Delta =0.5$ in Fig. \ref{fig3}. Clearly the values of Berry phases under
the RWA are close to those beyond the RWA in the weak coupling region. We
find that in RWA the berry phase for $\Psi _{0}$ is constantly zero, while
those for $\Psi _{1}^{2}$ and $\Psi _{1}^{3}$ swap their positions if the
detuning $\Delta $ reverses its sign. In strong coupling regime all states
acquire a much bigger Berry phase than in RWA, in some cases even exceeding $%
2\pi $.

In Fig. \ref{fig5}, we further compare the geometric phase for the evolution
of noneigenstate $\left\vert 10\right\rangle |0\rangle $ in these two cases.
It is shown that all geometric phases tend to a constant $\pi $ with the
increase of the coupling strength in the RWA and are symmetric for blue and
red detunings. The symmetry is broken beyond the RWA and we find an
anomalous sudden change for $\Delta =0.5\omega _{c}$ around $g\sim
0.265\omega _{c}$. We attribute this to the generic level anti-crossing in
the same parity subspace in Fig. \ref{fig3}(b). The geometric phase of the
noneigenstate $\left\vert 10\right\rangle |0\rangle $ here is the weighted
summation of the Berry phases (\ref{Bberry}) of all eigenstates in the
numerically truncated space. The transition between the two nearly
degenerate eigenstates at point (B) in Fig. \ref{fig3}(b) in the same parity
can not be neglected and the Berry phase in an adiabatic cyclic evolution
is ill defined. The geometric phase is no longer applicable whenever the
level anti-crossing in the same parity subspace occurs under some parameter
conditions.

\section{Conclusion}

To conclude, we have investigated theoretically the geometric curvature and
phase of the quantum Rabi model. When the wave function of the system undergoes
an adiabatic evolution along a closed curve in the parameter space, we derived the
analytical expressions of the Berry phase of the eigenstates under the RWA for the
single- and two-qubit systems. We introduced the Berry connection and curvature
to study the geometric properties of the Rabi model and it was found that the
curvatures for both single- and two-qubit models are identical to the example of
spin-1/2 particle. This idea is generalized to describe the vacuum-induced geometric
curvature when the system starts from an initial state with pure vacuum bosonic field,
which can be recognized as a statistical average of the magnetic field generated
simultaneously by two monopoles with opposite magnetic charges. The induced
geometric phase is related to the average photon number in a period, which is
different from the AA phase of the evolution. With the displaced Fock state technique
we managed to evaluate the Berry phase for eigenstates and the geometric phase
for pure vacuum initial state beyond the RWA. The anti-crossing points in the spectrum
invalidate the condition of the adiabatic theorem as the transition between the states
in the same parity space is not negligible - an anomalous sudden change occurs in the
geometric phase, which implies that the Berry phases of the eigenstates in an adiabatic
cyclic evolution are ill-defined near these nearly degenerate points.

When the two-qubit system undergoes a cyclic evolution, the time dependent eigenstates
will pick up Berry phases, which defines an adiabatic geometric phase gate acting on the
two-qubit basis states as well as their superpositions. As an example of this, Jones et al. \cite{Jones}
performed an experimental realization of a controlled phase shift gate by considering an ensemble
of spin half particles in a magnetic field, in which the fidelity that measures the precision of
the experimentally implemented gates with respect to an ideal one was higher than traditional
dynamical gates. Crucially, we have studied the Berry phase of the single- and two-qubit Rabi model,
where a single quantized mode of the field is considered and the phases depend on the detuning $\Delta$
and coupling strength $g$. In experiment, it is easy to control the detuning and the coupling
quite precisely. Therefore, the result here would be useful in the realization of geometric
phase gates in a circuit QED system on the basis of similar theories.
In particular, we should be more careful when considering two-qubit quantum logic
gates in many experimental implementations based on the geometric phase of the
model where employment of the RWA might impart incorrect results \cite%
{Leibfried, Kim}. Traditionally the validity of the RWA is guaranteed by the
energy conservation and the weak coupling in the system. This study shows
that the breakdown of the RWA also implies deviation of the geometric property of the
state evolution from the true strong coupling system.

\section*{Acknowledgements}

This work is supported by NSF of China under Grant Nos. 11234008 and
11474189, the National Basic Research Program of China (973 Program) under
Grant No. 2011CB921601, Program for Changjiang Scholars and Innovative
Research Team in University (PCSIRT)(No. IRT13076).

\end{document}